\documentclass[12pt]{article}

\usepackage{amsmath,amsfonts,amssymb, amsthm}
\usepackage{algorithmic}
\usepackage{algorithm}
\usepackage{array}
\usepackage[caption=false,font=normalsize,labelfont=sf,textfont=sf]{subfig}
\usepackage{textcomp}
\usepackage{stfloats}
\usepackage{url}
\usepackage{verbatim}
\usepackage{graphicx}
\usepackage{cite}

\usepackage{dsfont}

\usepackage{color}
\usepackage{cancel}
\usepackage{hyperref}

\newtheorem{proposition}{Proposition}
\newtheorem{example}{Example}

\newcommand{\be}{\begin{equation}}
\newcommand{\ee}{\end{equation}}

\newcommand{\al}{\alpha}
\newcommand{\tta}{\theta}
\newcommand{\ga}{\gamma}
\newcommand{\Ga}{\Gamma}

\newcommand{\bb}{\overline}

\newcommand{\E}{{\mathbb E}}
\newcommand{\cE}{{\cal E}}
\newcommand{\cH}{{\cal H}}
\newcommand{\N}{{\mathbb N}}

\newcommand{\nn}{\nonumber}
\newcommand{\q}{\quad}
\newcommand{\R}{{\mathbb R}}

\DeclareMathOperator*{\argmax}{argmax}
\DeclareMathOperator*{\Var}{{Var}}

\usepackage[margin=1in]{geometry}

 \definecolor{darkspringgreen}{rgb}{0.09, 0.45, 0.27} 
 \definecolor{darkgray}{rgb}{0.66, 0.66, 0.66}

\definecolor{DeepBlue}{RGB}{0,0,160} 
\newcommand{\Dblue}{\color{black}}

\begin{document}

\title{
Node Classification via Simplicial Interaction with Augmented Maximal Clique Selection
}

\author{
Eunho Koo\thanks{Department of Big Data Convergence, Chonnam National University, Gwangju 61186, Republic of Korea (kooeunho@jnu.ac.kr)}
\and
Tongseok Lim\thanks{Mitch Daniels School of Business, Purdue University, West Lafayette, Indiana 47907, USA (lim336@purdue.edu)}
}

\date{}

\maketitle





\begin{abstract}
Considering higher-order interactions allows for a more comprehensive understanding of network structures beyond simple pairwise connections. While leveraging all cliques in a network to handle higher-order interactions is intuitive, it often leads to computational inefficiencies due to overlapping information between higher-order and lower-order cliques. To address this issue, we propose an augmented maximal clique strategy. Although using only maximal cliques can reduce unnecessary overlap and provide a concise representation of the network, certain nodes may still appear in multiple maximal cliques, resulting in imbalanced training data. Therefore, our augmented maximal clique approach selectively includes some non-maximal cliques to mitigate the overrepresentation of specific nodes and promote more balanced learning across the network. Comparative analyses on synthetic networks and real-world citation datasets demonstrate that our method outperforms approaches based on pairwise interactions, all cliques, or only maximal cliques. Finally, by integrating this strategy into GNN-based semi-supervised learning, we establish a link between maximal clique-based methods and GNNs, showing that incorporating higher-order structures improves predictive accuracy. As a result, the augmented maximal clique strategy offers a computationally efficient and effective solution for higher-order network learning.
 
\end{abstract}


\section*{Keywords}
Node classification, semi-supervised, simplex, clique, maximal clique, node interaction, higher-order networks, hypergraph, probabilistic objective function.


\section{Introduction}

Systems have been analyzed using various methodologies, often represented as graphs, which involve identifying important nodes or hubs within the graph \cite{friedkin1991theoretical, gleich2015pagerank}. However, many real-world networks exhibit phenomena that cannot be fully explained by traditional graph-based approaches, which typically focus solely on pairwise interactions, i.e., interactions between two nodes connected by an edge. For instance, economic activities rely on agreements among buyers, sellers, and intermediaries \cite{bonacich2004hyper}, while relationships such as friendships \cite{easley2010networks} and large-scale online networks \cite{facchetti2011computing} involve interactions among multiple nodes. Therefore, there is a need to develop methods that extend beyond pairwise interactions to understand the mechanisms of complex networks. 

A notable approach to addressing the perspective of network analysis involves the application of higher-order network methodologies, which are based on a probability framework derived from the Stochastic Block Model (SBM) \cite{holland1983stochastic, ghoshdastidar2014consistency, lesieur2017statistical, kim2017community}. SBM, functioning as a generative model, enables the construction of graphs that exhibit community structures within distinct node communities. This model leverages a comprehensive set of nodes, partitioned into disjoint subsets covering all nodes, along with the probabilities of connections between nodes within each subset, serving as its foundational parameters. SBM’s utility extends to the exploration and elucidation of a network’s inherent structures, as well as to the facilitation of clustering endeavors \cite{lee2019review}. Meanwhile, in the domain of traditional supervised learning methodologies, the necessity for extensive labeled datasets for classification tasks poses a significant challenge due to the difficulty of acquiring such data. To address the labor-intensive process of data labeling, the field of semi-supervised learning has emerged as a promising alternative \cite{xiaojin2008semi}. This area is actively researched \cite{macskassy2007classification, baluja2008video, talukdar2008weakly}, typically involving the strategic selection of a minimal subset of nodes to serve as prior knowledge in node classification tasks. Empirical evidence suggests that even a limited amount of prior information can significantly enhance prediction performance \cite{ver2011statistical, ma2010semi, allahverdyan2010community, zhang2013community, zhang2013enhanced}.

Recent advancements have introduced methodologies to leverage the capabilities of higher-order interactions \cite{battiston2020networks, bick2023higher, kim2024higher}. Specifically, in the context of semi supervised node classification, \cite{koo2025node} suggests utilizing node interactions within each higher-order clique in the network. Throughout the optimization phase, the clique interaction-based optimization scheme in their work enforces a requirement for nodal entities within a given clique to exhibit similar probability distributions, based on the hypothesis that nodes with dense interconnections are likely to share similar distribution characteristics. Consequently, nodes situated within higher-order cliques incur a greater penalty for distributional diversity compared to their counterparts in lower-order cliques. This methodology has demonstrated superior performance in classification tasks over traditional non-hypergraph approaches, which rely solely on pairwise interactions. However, the comprehensive nature of higher-order cliques, which integrate multiple lower-order cliques, renders the utilization of all cliques for learning purposes inefficient. As a result, it appears clear that a meticulous choice of a subset of cliques is required.

In this regard, focusing on maximal cliques offers a promising direction for such a selective approach. A maximal clique, by definition, is a clique that cannot be expanded through the inclusion of an adjacent node, and therefore does not reside within a larger clique. We derive the expected number of maximal cliques in networks generated by the Planted Partition Model (PPM), a special case of SBM, and examine the fluctuation of this metric in response to increases in node number and nodal connection probabilities. Building upon this foundation, we propose an augmented maximal clique strategy that enhances the basic maximal clique approach by strategically incorporating additional non-maximal cliques to balance learning frequencies across nodes. While maximal cliques form the core of our approach, we augment them with selected non-maximal cliques to address the inherent frequency imbalance where some nodes appear more frequently in training cliques than others, thereby ensuring a more uniform distribution of learning opportunities across the network structure.

We conduct a comprehensive comparative analysis of prediction performance across four different strategies (pairwise interaction approach, all-clique approach, maximal clique strategy, and our proposed augmented maximal clique method) on synthetic graphs generated using the PPM in both balanced and imbalanced settings. In the experiments, we initialize the node probability distribution using discrete potential theory, leveraging an appropriate Dirichlet boundary value problem on graphs, whose solution can be efficiently computed 
\cite{bendito2003solving}. 

Furthermore, we propose a method to integrate the proposed strategy with GNN-based semi-supervised learning techniques to achieve additional performance gains. Most GNN models primarily focus on pairwise interactions between nodes and do not explicitly incorporate higher-order simplices into their objective functions. In contrast, the proposed strategy directly utilizes numerous high-order maximal cliques scattered throughout the network as key components of the objective function. By using the node classification prediction vectors from GNN models as the initial node probability distribution in our approach, we establish a connection between the two methods while effectively leveraging their respective strengths. We investigate how leveraging the strengths of both approaches results in enhanced performance and validate these improvements through experiments on real citation network datasets (Cora, CiteSeer, PubMed, and Coauthor-Physics.)

{\Dblue This work builds on our earlier study \cite{koo2025node}, which introduced a clique-based probabilistic objective function. Although the overall structure of the objective function remains unchanged, the current paper emphasizes enhancing computational efficiency and predictive accuracy by presenting a new strategy for systematically selecting the cliques included in the training objective.}

	Our contributions are outlined as follows:
	
• We derive the expected number of maximal cliques (as well as general cliques) in the PPM, allowing us to analyze the distribution of higher-order cliques as the number of nodes and connection probabilities increase. This provides quantitative evidence that the maximal clique-based classification strategy in this paper significantly reduces computational complexity compared to the all-clique strategy considered in \cite{koo2025node}.

• We empirically validate that the augmented maximal clique strategy outperforms other approaches (pairwise interaction-based, all-clique, and maximal clique-only strategies) through extensive experiments on both synthetic networks and real-world datasets.

This paper is structured as follows. Preliminaries are given in Section \ref{sec:preliminaries}. The augmented maximal clique strategy is described in Section \ref{model}. The experimental setup is detailed in Section \ref{setup}. The results are discussed in Section \ref{results}. The conclusion is presented in Section \ref{conclusion}.

\section{Preliminaries}\label{sec:preliminaries}

\subsection{Hypergraphs}

Conventional graph analysis primarily focuses on pairwise interactions between two nodes. The objective of this paper is to propose an efficient strategy capable of leveraging the hypergraph structure inherent in the underlying graph. A hypergraph is a generalization of the traditional graph in which an edge can contain more than two nodes \cite{bick2023higher}. We describe notation used for undirected hypergraphs in this study. An undirected graph $G=(V,E)$ consists of a node set $V=\{1,2,…,N\}$, and an edge set $E \subset \{(i,j) \, | \, i,j \in V,\, i \ne j\}$ where $(i,j) = (j,i)$. A hypergraph generalizes $E$ as $\cE \subset 2^V$ where $2^V$ denotes the power set of $V$, and we denote the hypergraph as $\cH=(V,\cE)$. For a positive integer $k$, we define $\sigma = \{n_1,n_2,…,n_k\} \in E$ as a $k$-clique if $(n_i,n_j)\in E$ for every $1 \le i<j \le k$. Let $K_k$ denote the set of all $k$-cliques in $G$, such that $K_1$, $K_2$, $K_3$, and $K_4$ represent the sets of nodes, edges, triangles, and tetrahedra in $G$, respectively. The collection of all $k$-cliques in $G$, denoted by $K = \bigcup_{k=1}^M K_k$, is referred to as the clique complex of the graph $G$, where $M$ represents the size of the largest clique in $G$. We interpret $\cE$ as a subset of $K$ in the context of a hypergraph.

\subsection{Planted Partition Model}\label{sec:PPM}

The stochastic block model originates from the field of social networks \cite{holland1983stochastic}. In this model, the node set is partitioned into disjoint nonempty subsets $C_i$ for $i=1,2,…,l$, such that $V= \bigcup_{i=1}^l C_i$ and $C_i \cap C_j = \emptyset$ if $i \ne j$. Additionally, there exists a symmetric $l \times l$ matrix of edge probabilities, where the connection probability between two nodes $u$ and $v$ is the $(i,j)$ component of the matrix for all $u \in C_i$ and $v \in C_j$. The Planted Partition Model (PPM) represents a special case of the stochastic block matrix, where the entries of the $l \times l$ matrix are constant $p$ on the diagonal and $q$ off the diagonal, with $p>q$. A key advantage of PPM is that it can incorporate realistic factors such as noise and sparsity while maintaining a clearly defined community structure.

\subsection{Node Classification Algorithm Based on Random Walk on Graphs}\label{sec:RW}

In semi-supervised node classification tasks, a widely used algorithm based on random walks on a graph operates as follows. An unbiased random walk transitions from node $i \in V$ to node $j \in V$ with probability $1/d$ if $(i,j) \in E$ (i.e., $i$ and $j$ are adjacent), where $d$ represents the degree of $i$ (the number of nodes adjacent to $i$). Given a node set $V=\{1,2,...,N\}$ and a label-index set $I=\{1,2,...,l\}$, each node corresponds to a label in $I$, where only a small proportion of labels are known. For $i \in I$ and $y \in V$, let $p_i (y)$ denote the probability that a random walk starting from an unlabeled node $y$ reaches an $i$-labeled node before reaching any other labeled node. If $\argmax_{i \in I}   p_i (y)=k$, the algorithm assigns the label $k$ to the node $y$. (If a node $y$ already has the label $k$, then $p_i (y)=1$ if and only if $i=k$.)

The task is to determine $p_i (y)$ for all $y \in V$ and $i \in I$. It is known that $p_i (y)$ can be derived as the solution to the following Dirichlet boundary value problem:
\begin{align}\label{Dirichlet}
Lp_i(x) &= 0 \ \text{ if } \ x \in F = (E_i \cup H_i)^c, \\
p_i(x) &= 1 \ \text{ if } \ x \in E_i, \nn \\
p_i(x) &= 0 \ \text{ if } \ x \in H_i, \nn
\end{align}
where $L=D-A$ is the graph Laplacian matrix, $D$ and $A$ are the degree and adjacency matrices of the given graph $G$, respectively \cite{lim2020hodge}. $E_i$ is the set of $i$-labeled nodes, $H_i$ is the set of labeled nodes excluding $i$-labeled nodes. 

Bendito, Carmona, and Encinas \cite{bendito2003solving} proposed a novel approach to solve the Dirichlet problem \eqref{Dirichlet} using equilibrium measures. For any decomposition $V=F \cup F^c$ where $F$ and $F^c$ are both non-empty, they proved the existence of a unique measure (function) $v$ on $V$ such that $Lv(x)=1$ (and $v(x)>0$) for all $x$ in $F$ and $Lv(x)=0$ (and $v(x)=0$) for all $x$ in $F^c$. This measure is called the equilibrium measure and denoted by $v^F$. For $V=F \cup F^c$ where $F$ and $F^c$ are the sets of unlabeled and labeled nodes, respectively, they showed that the solution $p_i$ to \eqref{Dirichlet} can be represented as:
\be\label{Dirichletsol}
p_i (x) = \sum_{z \in E_i} \frac{v^{ \{z\} \cup F}(x) - v^{F}(x)}{v^{ \{z\} \cup F}(z)}, \q x \in V.
\ee
Since $v^F$ can be obtained by solving a linear program, \eqref{Dirichletsol} offers an efficient method for solving the Dirichlet problem \eqref{Dirichlet}; see \cite{bendito2003solving} for details. We refer to the algorithm described above as \textsf{RW}, short for random walk. The \textsf{RW} method not only serves as a performance benchmark for hypergraph-based strategies, but also plays a crucial role in node probability initialization. This initial step lays the basis for the subsequent training of the objective function that we shall now describe.

\subsection{Clique-based Probabilistic Objective Function}

We employ the clique-based probabilistic objective function for hypergraphs as proposed in \cite{koo2025node}. Given a graph with a node set $V=\{1,2,...,N\}$ and a label set $I=\{1,2,...,l\}$, the probability distribution assigned to a node $j$ is denoted by $p^j=(p_1^j,p_2^j,…,p_l^j)$, where $p_i^j$ denotes the probability that node $j$ has label $i$, so that $\sum_{i=1}^l p_i^j = 1$ for all $j \in V$. Recall that $K_k$ denotes the set of $k$-cliques, and $M$ is the maximum possible value of $k$ in the graph. Define the permutation set with repetitions (denoted by $S_k$) as the set of ordered and repetition-allowed arrangements of $k$ elements in $I=\{1,...,l\}$, so that $|S_k| =l^k$. Then, the proposed objective function on hypergraph for node classification is given by
\begin{align}\label{objective}
J = \sum_{k=2}^M W_k \sum_{(n_1,...,n_k) \in K_k} \sum_{(m_1,...,m_k)=\tta \in  S_k} C_\tta p_{m_1}^{n_1}  p_{m_2}^{n_2}\dots  p_{m_k}^{n_k}
\end{align}
where $W_k>0$ are weight parameters, $C_\tta := {k \choose e_1,e_2,...,e_l} = \frac{k!}{e_1!e_2!\dots e_l!}$ where $e_i$ is the number of occurrences of the label $i$ in $\tta = (m_1,m_2,…,m_k )$ such that $\sum_{i=1}^l e_i=k$. Observe that $C_\tta$ is larger as the label frequencies $e_1,e_2,...,e_l$ are more uniform. Now the optimization problem we consider is:
\begin{align}\label{problem}
\text{Minimize } \, J \, \text{ over }  \, \Delta^N = \Delta_1 \times \Delta_2 \times \dots \times \Delta_N
\end{align}
where $\Delta_j = \{ p^j = (p_i^j)_{i \in I} \mid \sum_{i=1}^l p_i^j = 1,\, p_i^j \ge 0\}$ is the probability simplex in $\R^l$. Observe that the objective function $J$ assigns a higher penalty to cliques with a greater diversity of labels and vice versa, for each $k$-clique. The penalties are summed for all $k$-cliques, and then aggregated with weight $W_k$ for each $K_k$. This function is formulated under the hypothesis that nodes within higher-order cliques in the network tend to have similar labels. Consequently, the objective function is designed such that higher-order cliques exert greater pressure on their constituent nodes to adopt similar labels compared to lower-order cliques.

{\Dblue It is worth noting that the problem of finding all maximal cliques in a graph, known as Maximal Clique Enumeration (MCE), is a fundamental challenge in graph theory. The Bron-Kerbosch algorithm \cite{bron1973algorithm} is a pioneering backtracking method developed for this task. Subsequent research has focused on improving its worst-case time complexity, with Tomita's algorithm \cite{tomita2006worst}, which optimizes the pivot selection strategy, being a widely recognized and influential benchmark. While our work does not propose a new enumeration algorithm, we build upon the outputs of such standard methods to develop our higher-order node classification framework.}

\subsection{Literature Review on Graph Semi-Supervised Learning}

Graph Semi-Supervised Learning (GSSL) utilizes graph structures to improve learning performance when labeled data is scarce. It leverages the relationships between nodes in a graph, propagating label information from labeled to unlabeled nodes based on structural similarities. Methodologies for GSSL can be broadly categorized into graph regularization, matrix factorization-based methods, random walk-based methods, and Graph Neural Network (GNN)-based models.

Graph regularization methods leverage the manifold assumption to learn from the structure of the graph. One of the primary techniques in this category is Label Propagation (LP) \cite{xiaojin2002learning}, which propagates information from labeled nodes to unlabeled ones. This category includes Gaussian Random Fields (GRF) \cite{zhu2003semi} and Local and Global Consistency (LGC) \cite{zhou2003learning}. Additionally, general graph regularization techniques include manifold regularization \cite{belkin2006manifold, xu2010discriminative} that utilizes the graph Laplacian to smooth the relationships between nodes.

Matrix factorization-based methods learn low-dimensional representations of nodes by decomposing matrices that capture similarities between them. A well-known method is Laplacian eigenmaps \cite{belkin2001laplacian}, which utilize the graph Laplacian matrix for node embedding. Graph Factorization (GF) \cite{ahmed2013distributed} directly decomposes the adjacency matrix to obtain node representations, while GraRep \cite{cao2015grarep} factors multi-step transition probability matrices to create meaningful node embeddings. HOPE \cite{ou2016asymmetric} further preserves higher-order proximity by decomposing asymmetric matrices, maintaining structural information in a more refined way. On the other hand, random walk-based methods use stochastic sampling techniques to traverse the graph and capture structural information. DeepWalk \cite{perozzi2014deepwalk} is a widely used approach in this category, generating random walk sequences from the graph and training node embeddings using a Word2Vec-inspired model. Node2Vec \cite{grover2016node2vec} extends this idea by incorporating biased random walks that balance between breadth-first search and depth-first search, allowing it to capture more nuanced relationships. Planetoid \cite{yang2016revisiting} further integrates random walks with additional semi-supervised learning mechanisms to enhance prediction performance.

GNN-based models use Graph Neural Networks to extract rich node representations. Graph Convolutional Networks (GCN) \cite{kipf2017semisupervised} apply convolutional operations over graph structures to generate node embeddings. Graph Attention Networks (GAT) \cite{velikovi2018graph} introduce an attention mechanism that assigns different weights to neighbors, enabling more fine-grained feature learning. Simplifying Graph Convolutional networks (SGC) \cite{pmlr-v97-wu19e} removes non-linearities between GCN layers and collapses weight matrices, significantly reducing computational overhead while maintaining comparable performance. GraphSAGE \cite{hamilton2017inductive} improves scalability by sampling neighbor nodes instead of using the entire graph during training. Jumping Knowledge Networks (JKNet) \cite{xu2018representation} aggregate representations from multiple layers to capture hierarchical information and improve prediction accuracy.

{\Dblue Finally, recent studies have further advanced graph representation learning by exploring a variety of approaches to leverage complex graph structures. For instance, Motif-based GNNs (MGNN) explicitly incorporate predefined higher-order structures, including triangles (i.e., 3-cliques), to capture local topology beyond simple pairwise edges \cite{chen2023motif}. In contrast, other methodologies that do not rely on predefined structures have also been actively investigated. Probabilistic approaches, exemplified by SDMG \cite{zhusdmg}, implicitly capture complex relationships by learning the graph's overall distribution through diffusion models. Meanwhile, methods that take an adaptive approach, notably CSSE \cite{du2024customized}, utilize neural architecture search to discover and encode the most informative subgraph patterns for a given task, which offers an alternative to clique-based methodologies like ours.}

\section{Augmented Maximal Clique Approach}\label{model}
\begin{figure}[!t]
\centering
\includegraphics[width=\textwidth]{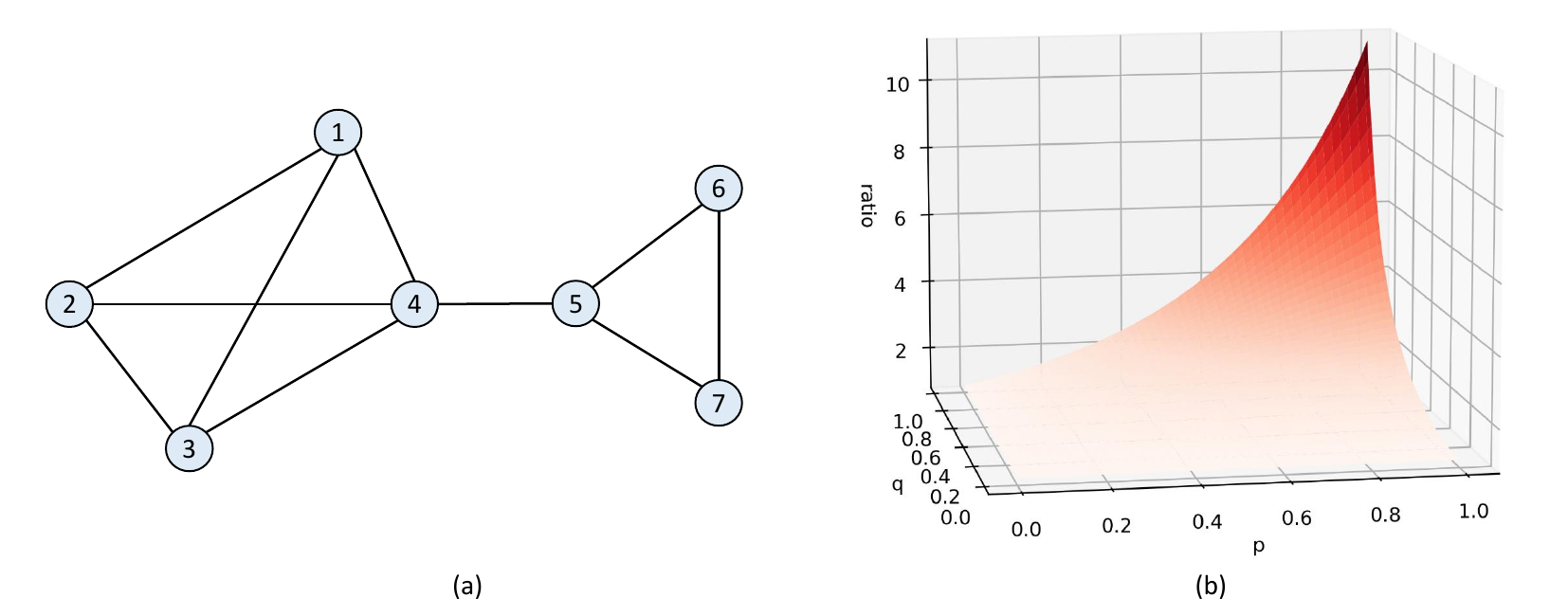}
\caption{In the graph in (a), there are 16 cliques: $K_2 = \{ (1,2), (1,3), (1,4), (2,3), (2,4), (3,4), (4,5), (5,6), (5,7), (6,7) \}$, $K_3 = \{(1,2,3), (1,2,4), (1,3,4), (2,3,4), (5,6,7) \}$, and $ K_4 = \{ (1,2,3,4) \}$. On the other hand, there are 3 maximal cliques: \(Q_2 = (4,5) \), \( Q_3 = (5,6,7) \), and $Q_4 = (1,2,3,4) \). An illustration of the ratio $\frac{\sum_{k=2}^4 \mathbb{E}[|K_k|] }{\sum_{k=2}^4 \mathbb{E}[|Q_k |]}$ in Example \ref{ex1} is presented in (b).}
\label{figure:intro}
\end{figure}

Recent research findings (referenced in \cite{tang2024hypergraph, lee2018higher} for instance) underscore that integrating higher-order cliques into the training process substantially improves node classification performance. This enhancement stems from the recognition that higher-order cliques capture a more intricate network of node relationships, thereby providing a richer context for learning algorithms. Specifically, the objective function \eqref{objective} is tailored to apply increased pressure on nodes within these complex cliques to adopt more homogeneous labels. However, addressing the minimization problem of $J$ on the domain $\Delta^N$ becomes exceedingly challenging due to the rapidly increasing number of total cliques in the graph. Furthermore, this approach suffers from a frequency imbalance problem, as nodes that appear in multiple cliques are trained more frequently, potentially biasing the learning process.

This paper is motivated by the question of how to effectively reduce the computational complexity associated with problem \eqref{problem} while preserving the enhanced prediction performance through higher-order clique interactions, and how to address the frequency imbalance issue where nodes appearing in multiple cliques are overrepresented during the training process. As a result, we propose the {\it augmented maximal clique strategy}, using an objective function that leverages maximal cliques along with select additional cliques, rather than relying on all cliques in the network. A maximal clique is a clique that is not a subset of a larger clique (Figure \ref{figure:intro}(a)). Let $Q$ $(Q_k)$ denote the set of maximal cliques (maximal $k$-cliques) in $G$. At first, we investigate the reduced objective function
\begin{align}\label{objective2}
J_Q = \sum_{k=2}^M W_k \sum_{(n_1,...,n_k) \in Q_k} \sum_{(m_1,...,m_k)=\tta \in  S_k} C_\tta p_{m_1}^{n_1} \dots  p_{m_k}^{n_k}.
\end{align}
As no clique subsumes other cliques appearing in $J_Q$, no clique interaction is counted twice in $J_Q$. To illustrate the computational reduction achieved using $J_Q$, we derive the expected number of all cliques and maximal cliques in a network generated by PPM. We begin with a simple example.

\begin{example}\label{ex1}
Let \( V=\{a,b,c,d\} \) and \( I=\{1,2\} \), so there are four nodes and two labels in the network. Let the nodes \( C_1 = \{a,b\} \) carry label \( 1 \), and the nodes \( C_2=\{c,d\} \) carry label \( 2 \). Let the graph \( G = (V,E) \) be generated by PPM with parameters \( p, q \) as in Section \ref{sec:PPM}. Then the expected number of edges in \( G \), denoted by \( \mathbb{E}[|K_2|] \), is easily computed as \( \mathbb{E}[|K_2|] = 2p+4q \). Similarly, the expected number of triangles \( \mathbb{E}[|K_3|] = {4 \choose 3} pq^2 = 4pq^2 \), and the expected number of tetrahedra \( \mathbb{E}[|K_4|] = p^2q^4 \). Hence, the expected number of cliques in \( G \) is given by \( \sum_{k=2}^4 \mathbb{E}[|K_k|] = 2p+4q + 4pq^2 + p^2q^4 \). 

To calculate the expected number of maximal cliques in $G$, consider the edge $(a,b)$ as an example. For $(a,b)$ to be a maximal clique in $G$, the condition is $\{(a,c), (b,c)\} \nsubseteq E$ and $\{(a,d), (b,d)\} \nsubseteq E$. That is, the edge $(a,b)$ should not be contained in a triangle in $G$. The probability of this occurring is $(1-q^2)(1-q^2) = (1-q^2)^2$. Thus, the probability of the existence of the edge $(a,b)$ with it being a maximal clique is $p(1-q^2)^2$. Similarly, the probability of the existence of the edge $(a,c)$ with it being a maximal clique is $q(1-pq)^2$. Therefore, the expected number of maximal edges is $2p(1-q^2)^2 + 4q(1-pq)^2$. For a triangle $(a,b,c)$ to be maximal, it requires $\{(a,d),(b,d),(c,d) \} \nsubseteq E$, with a probability of $1-pq^2$. Hence, the expected number of maximal triangles is $4pq^2(1-pq^2)$. Finally, the tetrahedron $(a,b,c,d)$ is maximal, with a probability of existence of $p^2q^4$ as before. Therefore, the expected number of maximal cliques in the PPM is given by $\sum_{k=2}^4 \mathbb{E}[|Q_k|] = 2p(1-q^2)^2 + 4q(1-pq)^2 + 4pq^2(1-pq^2) + p^2q^4$. Figure \ref{figure:intro}(b) illustrates the ratio $\frac{\sum_{k=2}^4 \mathbb{E}[|K_k|] }{\sum_{k=2}^4 \mathbb{E}[|Q_k |]}$. 
\end{example}

In Proposition \ref{cliqueformula} in Appendix \ref{appendix}, we shall derive detailed explicit formulas and derivations for the expected number of all cliques and maximal cliques for general $N,p,q$, enabling the estimation of the computational efficiency of a strategy using only maximal cliques compared to one using all cliques by evaluating the ratio of the number of cliques used.

The approach utilizing maximal cliques in \eqref{objective2} aims to eliminate redundant clique overlap learning and improves computational efficiency. However, it still suffers from a key limitation as \eqref{objective}, namely, the unevenly distributed in node participation frequency during training. In particular, hub nodes in the network are still likely to appear in multiple maximal cliques, leading to excessive learning for certain nodes. Meanwhile, some cliques participate very infrequently in training, such as edges that do not contain hub nodes within some maximal clique. The objective functions \eqref{objective}, \eqref{objective2} incorporate the probability vectors of nodes in each clique into the training process, ensuring that training is concentrated on nodes with high participation frequencies. Therefore, adjusting node participation frequencies to achieve a more even distribution is essential for improving node classification performance. 
The augmented maximal clique strategy to address this issue is as follows. First, we take the set of maximal cliques as the initial clique set. For all nodes included in maximal cliques, let the participation frequency be denoted as $\ga=(\ga_1,\ga_2,...,\ga_N)$ where $N$ is the number of nodes in network, and $\ga_i$ denotes the number of maximal cliques containing the $i$th node. We define the mean participation frequency as $\Ga =1/N \sum_{i=1}^N \ga_i$. Now, suppose we select a $k$-clique $\psi$ that is not a maximal clique (i.e., $\psi \in K \setminus Q$ and $|\psi|=k$) and add it to the set of maximal cliques. Let the updated participation frequency be $\ga'=(\ga_1',\ga_2',...,\ga_N')$ after this modification. We then compare the variance of $\ga$ and $\ga'$. For each $i=1,2,...,N$, let $\xi_i= \ga_i - \Ga$, then it follows that $\sum_{i=1}^N \xi_i =0$ and $\Var(\ga) = 1/N \sum_{i=1}^N (\ga_i - \Ga)^2 = 1/N \sum_{i=1}^N \xi_i^2$. Similarly, for $\ga_i'$ (which is $\ga_i +1$ if $i \in \psi$, $\ga_i$ otherwise), $\Ga' = 1/N (\sum_{i=1}^N \ga_i + k) = \Ga + k/N$ and $\xi_i' = \ga_i' - \Ga' = \begin{cases}
\xi_i + (1 - k/N), \ i \in \psi,\\
\xi_i - k/N, \ \text{otherwise,}
\end{cases}
$hence
\be\label{xi}
\sum_{i=1}^N \xi_i'^2 = \sum_{i=1}^N \xi_i^2 + 2\sum_{i \in \psi} \xi_i + \frac{k(N-k)}{N}.
\ee
If the sum of the last two term is negative, the variance of $\ga'$ becomes smaller than that of $\ga$. This means that by selecting $\psi$ such that $\sum_{i \in \psi} \xi_i < \tfrac{k(k-N)}{2N}$, the learning frequency of nodes can be made more evenly distributed. {\Dblue To systematically select these non-maximal cliques, we employ a greedy iterative approach. This process begins with the initial set composed of all maximal cliques and its corresponding variance of node participation frequencies. In each iteration, we greedily select the single non-maximal clique from the candidates that yields the greatest reduction in the current variance. Once a clique is added to the set, the node participation frequencies and variance are immediately updated, and this new state serves as the baseline for the selection in the next iteration. This process is repeated until a predefined budget (i.e., the maximum number of non-maximal cliques to be added) is exhausted or no clique can be found that further improves the variance. The set of cliques resulting from this procedure is defined as the augmented maximal clique set (denoted by $\bb{Q}$), and the algorithm is presented in Table \ref{table1}.} 
  \begin{table}[!h]
\centering
\includegraphics[width=5in]{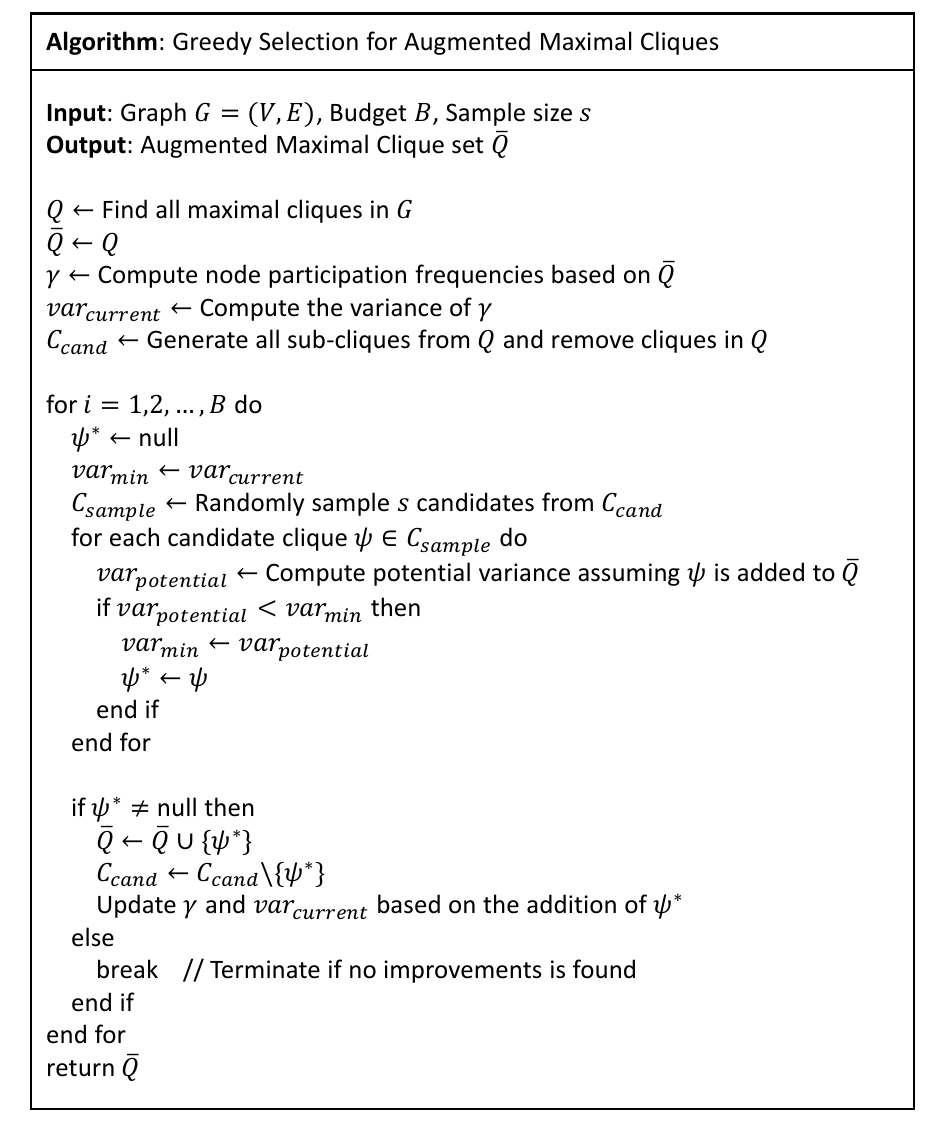}
\caption{{\Dblue Greedy algorithm for constructing the \textsf{Aug-MAX} set.}}
\label{table1}
\end{table}

Under this definition, the objective function \eqref{objective} becomes:
\begin{figure*}[!h]
\centering
\includegraphics[width=\textwidth]{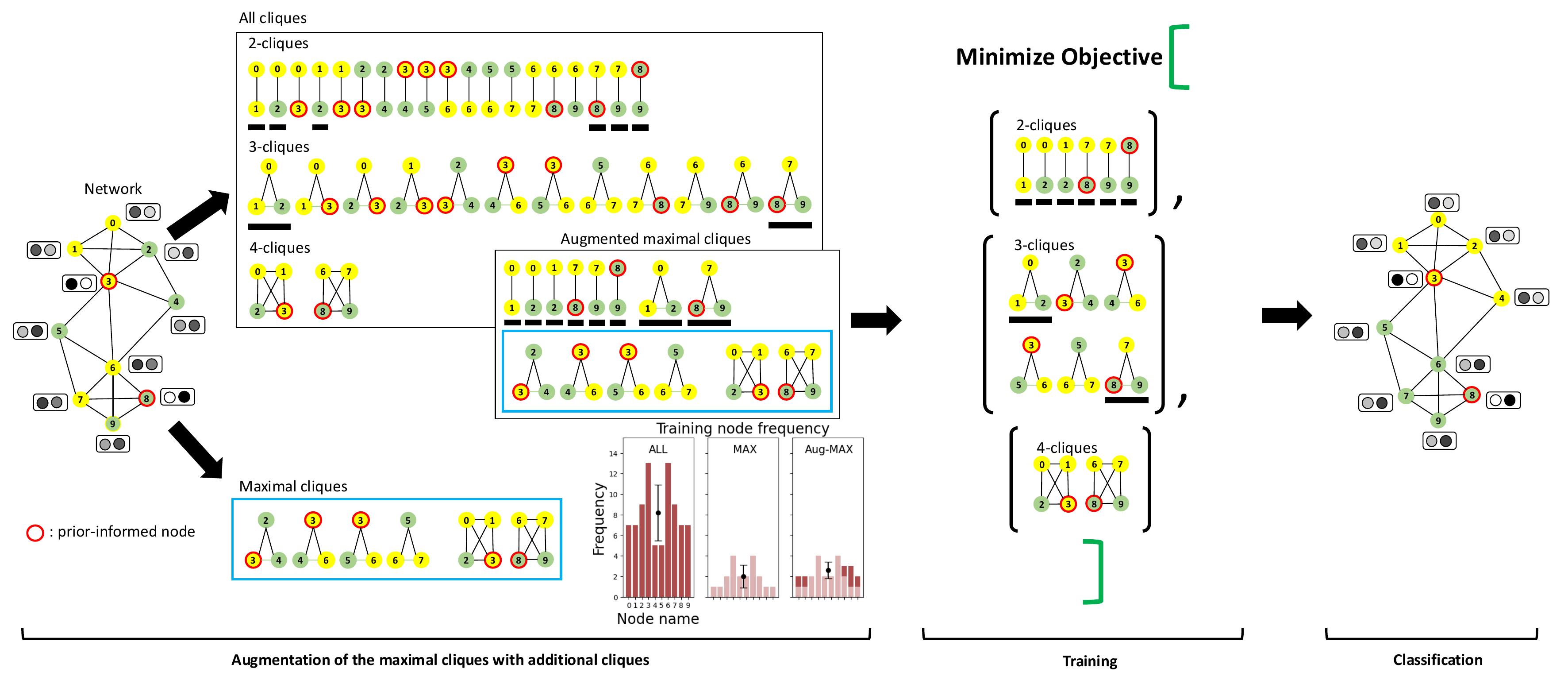}
\caption{Overview of the experimental pipeline based on the augmented maximal clique strategy. The model first selects cliques for learning by incorporating certain non-maximal cliques into the set of maximal cliques (left), then performs learning on the resulting higher-order network structure (middle), and finally conducts classification (right).}
\label{figure:process}
\end{figure*}
\begin{align}\label{objective3}
J_{\bb Q}= \sum_{k=2}^M W_k \sum_{(n_1,...,n_k) \in {\bb Q}_k} \sum_{(m_1,...,m_k)=\tta \in  S_k} C_\tta p_{m_1}^{n_1} \dots  p_{m_k}^{n_k}.
\end{align}
The objective \eqref{objective3} allows overlapping cliques to be learned, in contrast to the maximal clique strategy \eqref{objective2}. Figure \ref{figure:process} illustrates the concept of the augmented maximal clique strategy.

\begin{figure}[!t]
\centering
\includegraphics[width=3in]{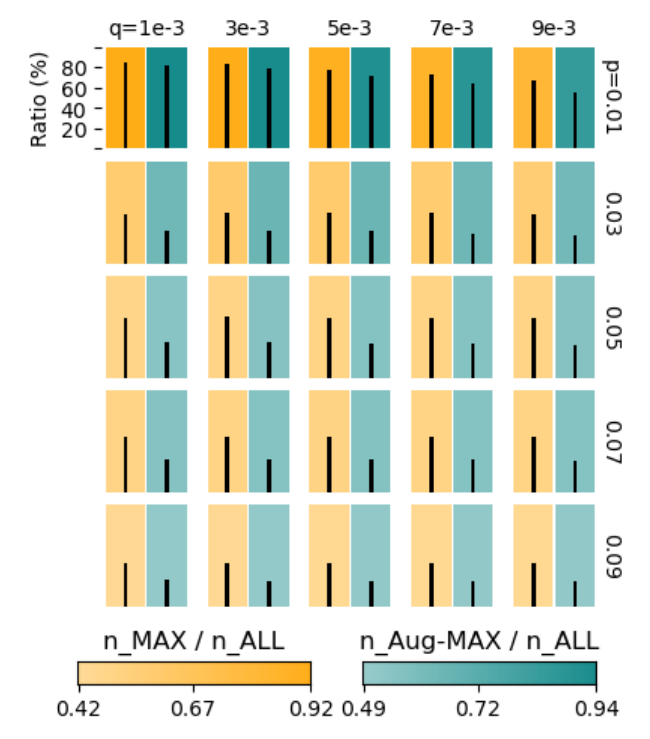}
\caption{Relative proportion of maximal (orange) and augmented maximal (teal) cliques in the PPM model with 5 labels (500 nodes per label) depending on the values of $p$ (intra-connection probability) and $q$ (inter-connection probability). Black bars represent the variability (standard deviation) in node participation frequencies during training.}
\label{figure:eff}
\end{figure}

The computational complexity of the objective \eqref{objective} associated to the problem \eqref{problem} can be estimated as $O(\sum_{k=2}^M |K_k | l^k )$, where $l$ is the number of labels. In the maximal clique approach, the complexity would be $O(\sum_{k=2}^M |Q_k | l^k)$.
This indicates that the efficiency of the maximal clique strategy stems from the reduction in the number of cliques used for training. Notably, Figure \ref{figure:eff} shows that the proportion of maximal and augmented maximal cliques relative to the total number of cliques is comparable, and that the augmented maximal clique strategy significantly reduces the variance in node participation frequencies used for training. 

\section{Experimental setup}\label{setup}

To assess the performance of the proposed strategy, this study considers balanced networks and imbalanced networks generated by the PPM. The balanced graph consists of five labels $[100, 100, 100, 100, 100]$, meaning that each label corresponds to $100$ nodes (i.e., $|I|=5$ and $|V|=N=500$). The intra-connection probability (connection probability between nodes with the same label) $p$ ranges from $0.01$ to $0.1$, while the inter-connection probability (connection probability between nodes with different labels) $q$ ranges from $0.001$ to $0.01$. The imbalanced graph consists of six labels $[150, 150, 50, 50, 50, 50]$, resulting in $|I|=6$ and $|V|=N=500$. Finally, since this study is based on semi-supervised learning, the prior-informed ratio (the proportion of nodes whose labels are known) $r$ ranges from $0.01$ to $0.1$. 

The objective function corresponding to the experiment utilizing only edges ($2$-cliques) in the graph is expressed as
\begin{align}
J_{K_2} = \sum_{(n_1, n_2) \in K_2} \sum_{(m_1,m_2) =\tta \in I^2} C_\tta p_{m_1}^{n_1}  p_{m_2}^{n_2} \nn
\end{align}
where the experimental results obtained using $J_{K_2}$ are denoted as \textsf{PI} (stands for pairwise interactions). In addition to \textsf{PI}, we define experimental results obtained using the objective \eqref{objective} (and solving \eqref{problem}) as \textsf{ALL} (stands for all cliques), those obtained using \eqref{objective2} as \textsf{MAX} (stands for maximal cliques), and those obtained using \eqref{objective3} as \textsf{Aug-MAX} (stands for augmented maximal cliques). We initialize the node probability distribution for these four strategies using \textsf{RW} in Section \ref{sec:RW}. Furthermore, we report the accuracy gain of \textsf{ALL}, \textsf{MAX}, and \textsf{Aug-MAX} to \textsf{PI} and perform a comparative analysis of their accuracy performances. The accuracy gain is defined as $y/x-1$ where $x$ and $y$ represent the accuracy obtained by \textsf{PI} and the higher-order strategies (\textsf{ALL}, \textsf{MAX}, and \textsf{Aug-MAX}), respectively.



{\Dblue 
For the real dataset experiments, we employ a transductive semi-supervised learning setting following the methodology in \cite{yang2016revisiting, velikovi2018graph, sen2008collective}. In the citation datasets Cora, Citeseer, and Pubmed, we adopt the widely used Planetoid split, where 20 nodes per class are labeled for training, 500 nodes are used for validation, and 1,000 nodes are used for testing, while the remaining nodes are unlabeled but included in the graph during training. In contrast, for the Coauthor-Physics dataset, we follow a different splitting strategy: approximately 60\% of the nodes are used as the candidate pool for training, from which 20 nodes per class are sampled, 20\% of the nodes are reserved for validation with up to 100 nodes per class, and the remaining 20\% of the nodes are used for testing. We evaluate our method on four benchmark datasets under these settings. The Cora dataset consists of 2,708 nodes, 5,429 undirected citation edges, and 7 labels. The Citeseer dataset contains 3,327 nodes, 4,732 edges, and 6 labels. The Pubmed dataset comprises 19,717 nodes, 44,338 edges, and 3 labels. In all three citation datasets, nodes represent papers and edges represent undirected citation relationships. The Coauthor-Physics dataset contains 34,493 nodes and 495,924 edges, where nodes represent authors, edges denote coauthorship, and labels correspond to each author’s research field.

 We use a learning rate of 0.1, run 20 epochs, and set the default value of the weight hyperparameter $W_k$ to 1 in the objective function \eqref{objective3} for all experiments. The Adam optimizer \cite{KingmaB14} is employed for optimization. The implementation algorithm is available at \url{https://github.com/kooeunho/HOI-aug-max}.
}

\section{Results}\label{results}

In this section, we evaluate the node classification performance of the higher-order network-based objective function on balanced data (Section \ref{balanced}) and imbalanced data (Section \ref{imbalanced}) generated using PPM. In both sections, we compare the performance of the proposed objective function applied to 2-clique (edge) experiments (\textsf{PI}) and those incorporating higher-order cliques (\textsf{ALL}, \textsf{MAX}, and \textsf{Aug-MAX}). Additionally, we assess the accuracy gain of higher-order clique experiments relative to \textsf{PI}. Furthermore, we present and discuss experiments where higher-order clique structures are combined with various GNN-based semi-supervised node classification methods, leading to additional performance improvements (Section \ref{GNN}). In Section \ref{Extended Evaluation}, we provide an extended evaluation, and in Section \ref{summary}, we summarize and discuss the results.

\begin{figure*}[!t]
\centering
\includegraphics[width=\textwidth]{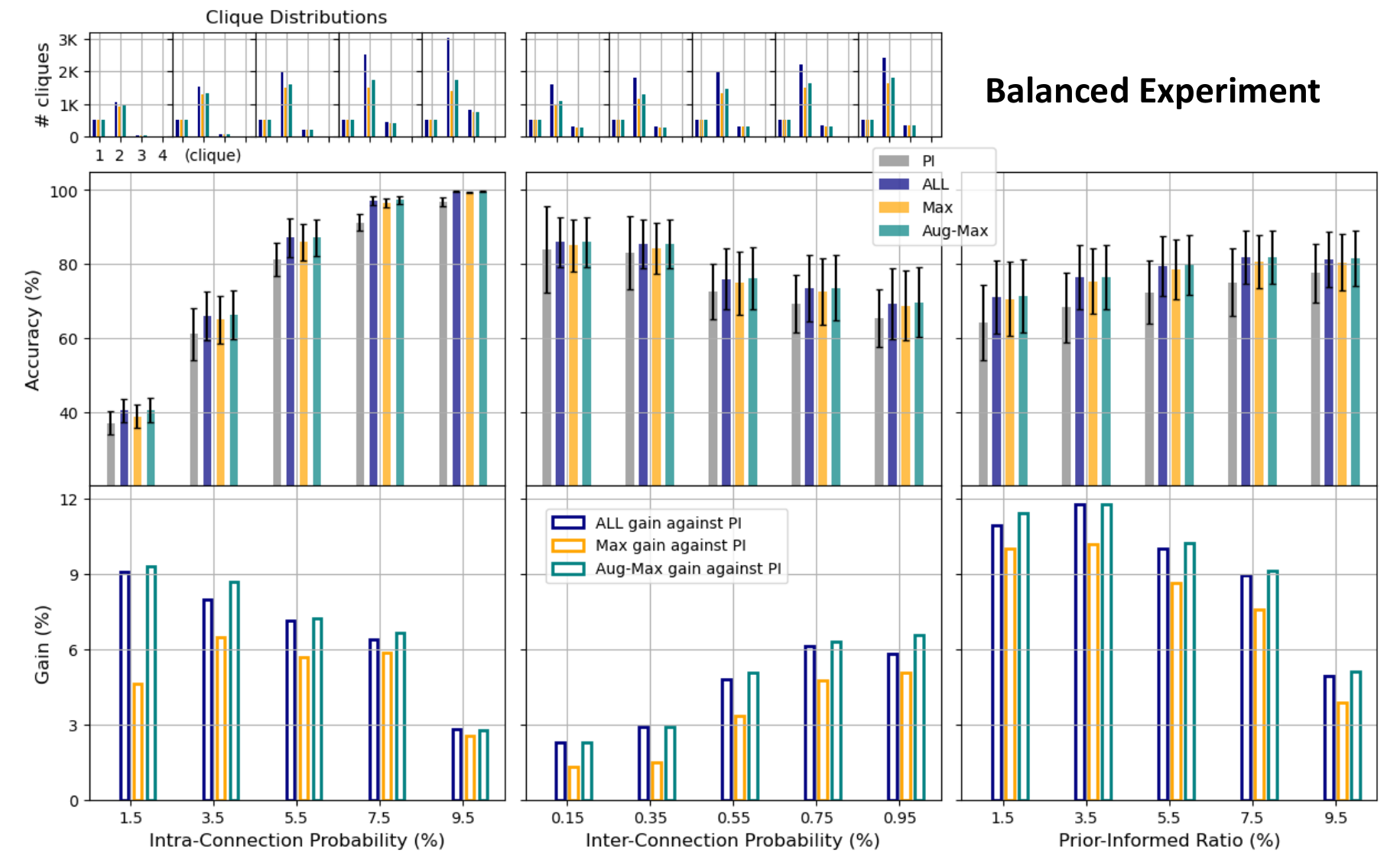}
\caption{Experimental results in the balanced setting [100, 100, 100, 100, 100]. The first row shows the clique distribution over the range of $p$ and $q$ (intra- and inter-class connection probabilities in PPM, respectively). The second row presents the classification accuracy for \textsf{PI} (black), \textsf{ALL} (navy), \textsf{MAX} (orange), and \textsf{Aug-MAX} (teal) across different ranges of $p$, $q$, and $r$ (prior-informed ratio). The third row illustrates the accuracy gains of higher-order clique-based approaches compared to \textsf{PI} over the same parameter ranges.}
\label{figure:bal}
\end{figure*}

\subsection{Balanced Experiment}\label{balanced}

\begin{figure*}[!t]
\centering
\includegraphics[width=\textwidth]{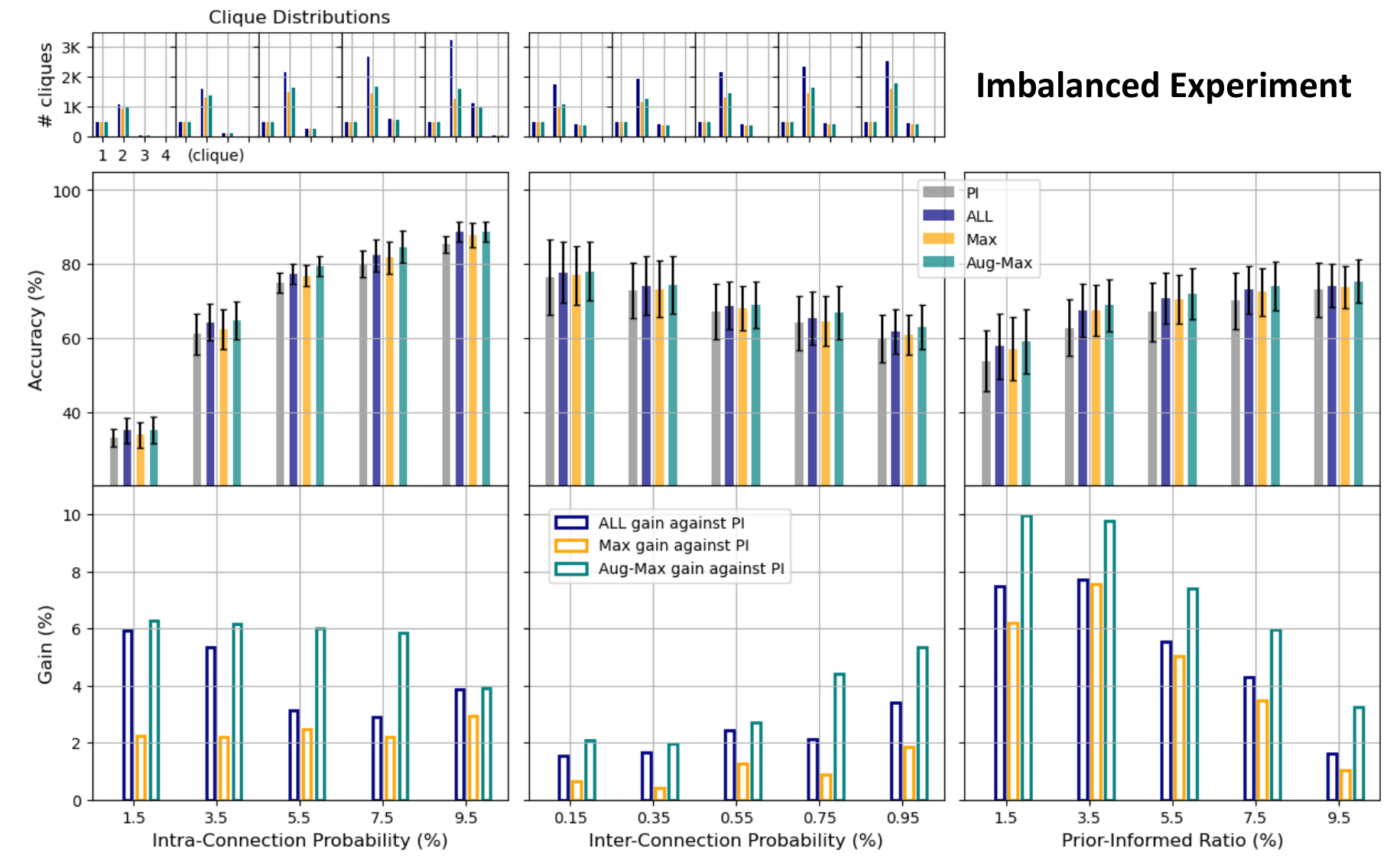}
\caption{Experimental results in the imbalanced setting [150, 150, 50, 50, 50, 50]. The first row shows the clique distribution over the range of $p$ and $q$ (intra- and inter-class connection probabilities in PPM, respectively). The second row presents the classification accuracy for \textsf{PI} (black), \textsf{ALL} (navy), \textsf{MAX} (orange), and \textsf{Aug-MAX} (teal) across different ranges of $p$, $q$, and $r$ (prior-informed ratio). The third row illustrates the accuracy gains of higher-order clique-based approaches compared to \textsf{PI} over the same parameter ranges.
}
\label{figure:imb}
\end{figure*}

We conducted experiments by varying the intra-class connection probability $p$ within the range [0.01, 0.1], the inter-class connection probability $q$ within [0.001, 0.01], and the prior information ratio $r$ within [0.01, 0.1]. For each combination of $p,q,r$, we measured the mean and standard deviation of node participation frequency in training, as well as the mean number of cliques used for training. These measurements were performed across all four experimental settings: \textsf{PI}, \textsf{ALL}, \textsf{MAX}, and \textsf{Aug-MAX}. The mean node participation frequencies across all $p,q,r$ combinations were 7.49, 9.67, 6.54, and 7.07 for \textsf{PI}, \textsf{ALL}, \textsf{MAX}, and \textsf{Aug-MAX}, respectively. The corresponding standard deviations were 2.68, 4.54, 2.35, and 2.19, while the mean number of cliques used for training was 1873, 2231, 1477, and 1609, respectively. Figure \ref{figure:bal} illustrates a comparative analysis of each strategy’s performance and the accuracy gain of higher-order clique strategies over \textsf{PI}.

Among the four strategies, \textsf{MAX} exhibits the lowest mean node participation frequency and the smallest mean number of cliques used (even compared to \textsf{PI}), indicating the highest computational efficiency. Compared to \textsf{MAX}, \textsf{Aug-MAX} shows slightly lower efficiency but yields performance gains by reducing the standard deviation of node participation frequency, thereby flattening its distribution across the network. Notably, \textsf{Aug-MAX} reduces the total number of training cliques by 28\% compared to \textsf{ALL}, while maintaining an average accuracy gain of 0.20\% across all $p,q,r$ configurations. 

Additionally, the performance gains of higher-order clique approaches (\textsf{ALL}, \textsf{MAX}, and \textsf{Aug-MAX}) over \textsf{PI} become more pronounced in challenging scenarios, i.e., when the intra-class connection probability was low, the inter-class connection probability was high, and the prior information ratio was low (Figure \ref{figure:bal}). This can be attributed to the ability of higher-order structures to capture richer information beyond simple one-hop (edge) relationships, such as multi-node co-occurrence patterns and highly connected substructures.

\subsection{Imbalanced Experiment}\label{imbalanced}

The imbalanced experiment is conducted under the same $p,q,r$ range settings as the balanced experiment. The mean node participation frequencies across all $p,q,r$ combinations are 8.05, 10.96, 7.02, and 7.66 for \textsf{PI}, \textsf{ALL}, \textsf{MAX}, and \textsf{Aug-MAX}, respectively. The corresponding standard deviations are 3.56, 6.07, 3.23, and 2.99, while the mean number of cliques used for training is 2012, 2491, 1545, and 1704, respectively. Figure \ref{figure:imb} illustrates a comparative analysis of each strategy’s performance and the accuracy gain of higher-order clique strategies (\textsf{ALL}, \textsf{MAX}, and \textsf{Aug-MAX}) over \textsf{PI}.

The key characteristics observed in the balanced experiment are also present in the imbalanced experiment. That is, compared to \textsf{MAX}, \textsf{Aug-MAX} shows a slight increase in the mean node participation frequency and the total number of cliques used but achieves improved accuracy by reducing the standard deviation of node participation frequency. Furthermore, \textsf{Aug-MAX} reduces the total number of training cliques by 32\% compared to \textsf{ALL}, while achieving an average accuracy gain of 1.42\% across all $p,q,r$ configurations. Additionally, similar to the balanced experiment, the performance gains of higher-order clique approaches (\textsf{ALL}, \textsf{MAX}, and \textsf{Aug-MAX}) over \textsf{PI} remain more pronounced in challenging scenarios.

\begin{figure*}[!t]
\centering
\includegraphics[width=\textwidth]{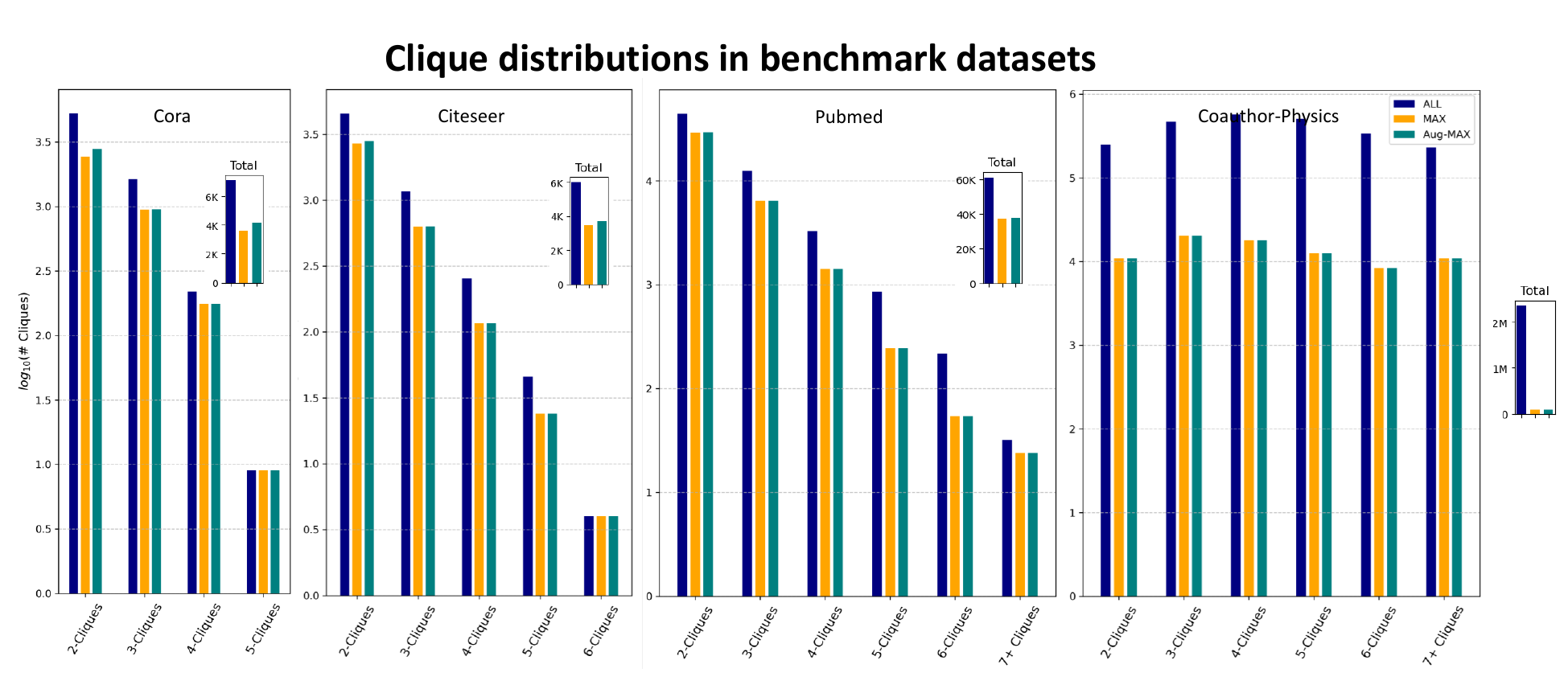}
\caption{
{\Dblue Clique distributions in the Cora, Citeseer, Pubmed, and Coauthor-Physics datasets. Navy, orange, and teal correspond to \textsf{ALL}, \textsf{MAX}, and \textsf{Aug-MAX}, respectively.}
}
\label{figure:CCP}
\end{figure*}

\subsection{Incorporating the proposed strategy into GNN}\label{GNN}
 
This section examines a strategy that enhances performance by integrating the proposed objective function with existing GNN-based node classification methods. Traditional GNN-based approaches learn node embedding representations through various techniques, such as adjusting the exploration and return variables of random walks \cite{grover2016node2vec} or employing attention mechanisms \cite{velikovi2018graph}. After obtaining node embeddings, these methods typically use a simple neural network structure or a linear transformation to derive the final probability distribution of node labels.

To effectively combine the higher-order connectivity-based objective function with GNN methods, we conduct experiments using the \textsf{ALL}, \textsf{MAX}, and \textsf{Aug-MAX} strategies, initializing them with the node label probability distributions produced by the baseline GNN models. Many GNN-based models do not directly incorporate higher-order structures during training. By leveraging this integration, the GNN architecture can incorporate various higher-order structures obtained from the \textsf{Aug-MAX} within the network based on the learned node probability distributions, ultimately leading to improved classification performance.

{\Dblue 
In the real dataset experiments, we evaluate models that integrate GNNs with higher order clique-based strategies using the Cora, Citeseer, Pubmed, and Coauthor-Physics datasets. Figure \ref{figure:CCP} illustrates the clique distribution and the number of cliques used with respect to \textsf{ALL}, \textsf{MAX}, and \textsf{Aug-MAX} in each dataset. Although budgets of 1000, 1000, and 3000 are allocated for Cora, Citeseer, and Pubmed, respectively, the number of cliques actually added to the maximal clique set to construct the \textsf{Aug-MAX} set is only 356, 111, and 191. This indicates that the number of candidate cliques capable of reducing the variance in node participation frequency is limited to these amounts. For Coauthor-Physics, the budget is set to 10,000, and the number of added cliques reaches this full amount, which is considered a sufficient augmentation as the rate of variance reduction diminishes significantly.

In the Cora experiments, the mean and standard deviation of node participation frequencies are 3.90, 6.05, 3.12, and 3.39, and 5.23, 10.43, 4.75, and 4.70 for \textsf{PI}, \textsf{ALL}, \textsf{MAX}, and \textsf{Aug-MAX}, respectively. For Citeseer, these values are 2.74, 4.17, 2.37, and 2.44, and 3.38, 9.31, 3.18, and 3.16, respectively. For Pubmed, they are 4.50, 7.36, 4.31, and 4.33, and 7.43, 23.49, 7.95, and 7.94, respectively. For Coauthor-Physics, 14.38, 303.02, 10.23, and 11.20, and 15.57, 935.46, 21.85, and 21.61, respectively. 

This study addresses the transductive learning-based semi-supervised node classification problem, using GAT \cite{velikovi2018graph}, GCN \cite{kipf2017semisupervised}, SGC \cite{pmlr-v97-wu19e}, Planetoid \cite{yang2016revisiting}, MGNN \cite{chen2023motif}, SDMG \cite{zhusdmg}, and CSSE \cite{du2024customized} as benchmark algorithms. These models generate embedding vectors using the node feature dimensions of the four datasets, which are 1433 (Cora), 3703 (Citeseer), 500 (Pubmed), and 8415 (Coauthor Physics), respectively. As a result of the experimental setup in Section \ref{setup}, the prior information for training and validation consists of 640 nodes for Cora, 629 for Citeseer, 560 for Pubmed, and 600 for Coauthor-Physics. We summarize the results in Table \ref{table2}.
}
  \begin{table}[!h]
\centering
\includegraphics[width=6in]{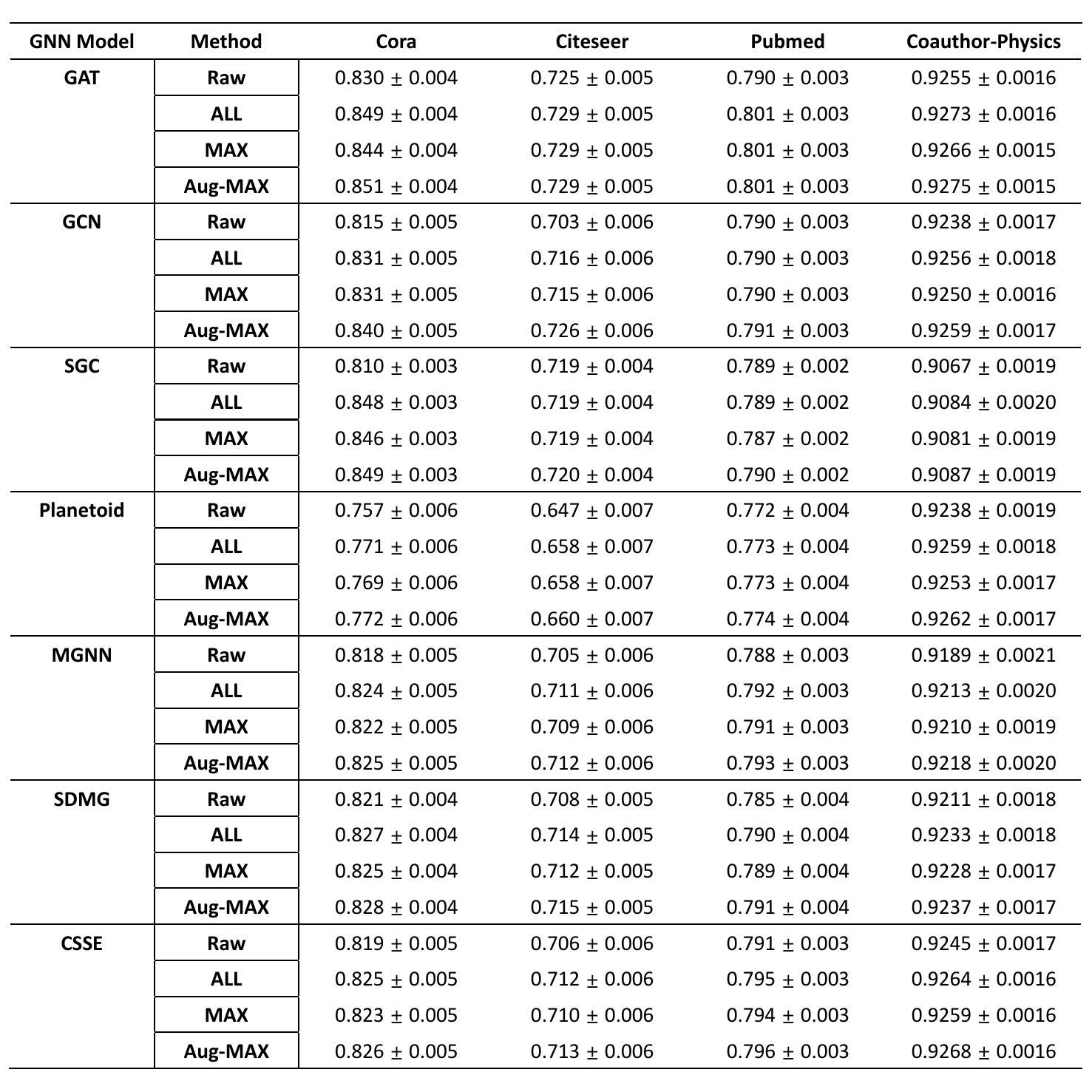}
\caption{Summary of accuracy results for benchmark datasets.}
\label{table2}
\end{table}
{\Dblue 
\subsection{Extended Evaluation}\label{Extended Evaluation}
\subsubsection{Weight Parameter $W_k$}
The objective function of our study includes the weight hyperparameter $W_k$ for each clique order $k$. While the main experiments in this paper employ a default strategy of uniform weighting ($W_k=1$), which assigns equal importance to cliques of all orders, we additionally analyze the impact of a linear weighting strategy (that is, $W_k=k$) that gives greater importance to higher-order cliques. 
On the synthetic datasets, our experimental results indicate that the linear weighting strategy yields a modest performance gain overall compared to the default uniform weighting setting. On average, we observe an accuracy improvement of 0.241\% in the balanced experiments and 0.947\% in the imbalanced experiments. More specifically, in the balanced experimental setting, the linear weighting leads to performance improvements of 0.182\%, 0.226\%, and 0.315\% for the \textsf{ALL}, \textsf{MAX}, and \textsf{Aug-MAX} strategies, respectively. This trend is more pronounced in the imbalanced setting, where performance gains of 0.784\%, 1.094\%, and 0.963\% are achieved for the respective strategies. This suggests that assigning higher weights to higher-order interactions can contribute to more effective learning of the network's complex structural information, particularly in scenarios with imbalanced class distributions. The performance advantage of the linear $W_k$ setting also extends to the GNN experiments, as presented in Table \ref{table3}. 
  \begin{table}[!h]
\centering
\includegraphics[width=4in]{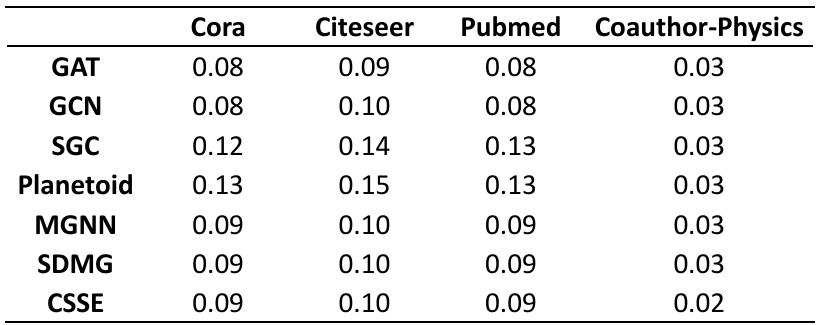}
\caption{{\Dblue Average accuracy gains (\%) of \textsf{ALL}, \textsf{MAX}, and \textsf{Aug-MAX} under the linear weighting scheme ($W_k=k$) compared to the uniform weighting scheme ($W_k=1$) across benchmark datasets.}}
\label{table3}
\end{table}
Averaged across the GNN models, the accuracy gains for the \textsf{ALL} strategy are 0.09\%, 0.10\%, 0.09\%, and 0.03\% on the Cora, CiteSeer, PubMed, and Coauthor-Physics datasets, respectively. The corresponding gains are 0.07\%, 0.08\%, 0.07\%, and 0.02\% for the \textsf{MAX}, and 0.12\%, 0.13\%, 0.11\%, and 0.03\% for the \textsf{Aug-MAX} strategy.

\subsubsection{Runtime Evaluation} 
To evaluate computational efficiency, the following training times are measured on the Coauthor-Physics dataset, with the augmentation budget set to 10,000. The baseline models require 55s (GAT), 15s (GCN), 35s (SGC), 20s (Planetoid), 45s (MGNN), 210s (SDMG), and 45s (CSSE). Using the initialized probability distribution from these GNNs, the training times for our higher-order strategies are approximately 130s for \textsf{ALL}, 10s for \textsf{MAX}, and 30s for \textsf{Aug-MAX}. It is important to note that this comparison excludes the one-time pre-processing cost, which consists of generating all cliques and maximal cliques (715s), constructing the \textsf{Aug-MAX} set (an additional 185s), and pre-computing the higher-order coefficients ($C_\tta$ in eqn. \eqref{objective3}) (45s). All experiments are conducted on a single GPU deep learning server configured with two Intel Xeon Gold 6426Y 16-core CPUs, 256GB of DDR5 RAM, and two NVIDIA L4 24GB GPUs. The software environment is built on the latest version of CentOS Linux, utilizing Python, PyTorch, and the NVIDIA CUDA Toolkit.

\subsubsection{Friedman Test}
We conduct a non-parametric statistical analysis across four real-world datasets (Cora, Citeseer, Pubmed, and Coauthor-Physics). We employ the Friedman test, a standard procedure for comparing multiple algorithms over multiple datasets. The performance of the 28 total models (\textsf{Raw}, \textsf{ALL}, \textsf{MAX}, and \textsf{Aug-MAX} corresponding to each of the 7 GNNs) is ranked separately for each of the four datasets. The null hypothesis (H0) is that all models perform equally, and thus their average ranks are statistically indistinguishable. Our analysis yields a Friedman chi-squared statistic of 658.32, with a corresponding $p$-value of $1.8 \times 10^{-121}$. As this $p$-value is substantially lower than the standard significance level ($\al=0.05$), we strongly reject the null hypothesis. 
Subsequently, we proceed with the Nemenyi post-hoc test. The Nemenyi test compares the average ranks of the models against a critical difference. If the difference in average ranks between two models is greater than the critical difference, their performance is considered to be statistically different (in our experiment with 28 models and 4 datasets, the critical difference at a significance level of $\al=0.05$ is 6.88). Table \ref{table4} presents the average ranks for the top 15 models.
  \begin{table}[!h]
\centering
\includegraphics[width=6in]{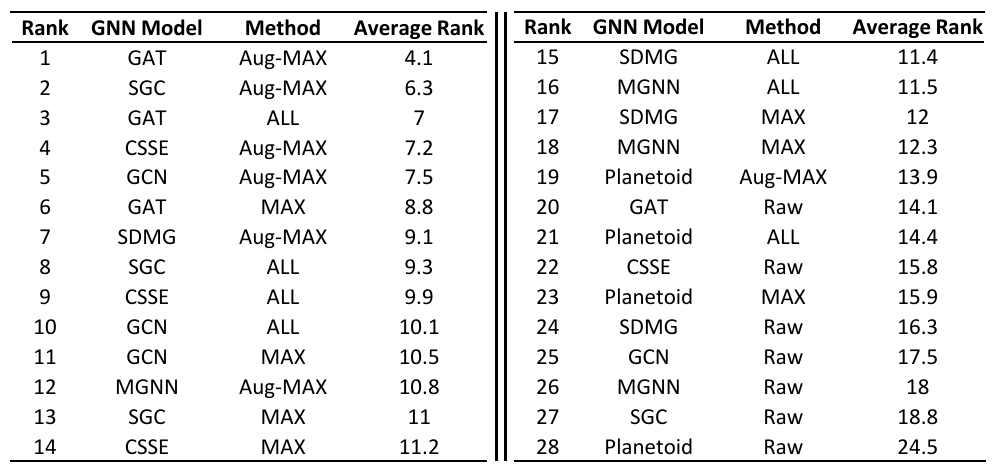}
\caption{{\Dblue Average ranks of all models based on the Friedman test. At significance level $\al=0.05$, the critical difference for the Nemenyi post-hoc test is 6.88.}}
\label{table4}
\end{table}
It is observed that for every GNN architecture, the \textsf{Aug-MAX} method consistently achieves a better rank than its corresponding Raw GNN baseline. This demonstrates the universal effectiveness of our proposed strategy. For instance, \textsf{GAT-Aug-MAX} (average rank 4.1) is statistically superior to \textsf{GCN-Raw} (average rank 17.5), as the rank difference of 13.4 is greater than the Critical Difference of 6.88. Additionally, the Nemenyi test identifies a top-performing group including \textsf{GAT-Aug-MAX}, \textsf{SGC-Aug-MAX}, \textsf{GAT-ALL}, \textsf{CSSE-Aug-MAX}, and \textsf{GCN-Aug-MAX}, within which the performance differences are not statistically significant. Nevetheless, the performance of this group is statistically superior to that of the lower-ranked models.

}

\subsection{Summary and Discussion}\label{summary}

 In this paper, we evaluate the proposed \textsf{Aug-MAX} strategy on balanced and imbalanced networks generated by PPM, as well as on real citation networks. The key findings and discussions derived from these experiments are as follows.
 
First, both \textsf{Aug-MAX} and \textsf{MAX} strategies utilize most of the higher-order cliques employed in \textsf{ALL}. More specifically, for $k \ge 2$, the ratio $|Q_k | / |K_k |$ (which is less than or equal to 1) approaches 1 as $k$ increases. This trend is observed in the upper sections of Figures \ref{figure:bal}, \ref{figure:imb} and \ref{figure:CCP}. This study employs objective functions \eqref{objective}, \eqref{objective2} and \eqref{objective3} that incorporate both lower-order and higher-order clique structures. We observe that, for large $k$, $Q_k$ and $K_k$ become nearly identical, implying that the three higher-order network-based learning strategies, \textsf{ALL}, \textsf{MAX}, and \textsf{Aug-MAX}, achieve nearly the same level of learning on higher-order cliques, which encapsulate hierarchical and collective properties of the network. As a result, despite the imbalance in the number of lower-order cliques used in training, their overall performance remains comparable.

Second, using all cliques in the network for training, which was used in the previous study \cite{koo2025node}, is the most intuitive approach in clique-based network analysis, but it may not be an optimal choice. In both synthetic networks and real networks used in this study, the standard deviation of node participation frequency in \textsf{ALL} is significantly higher than that in \textsf{MAX} and \textsf{Aug-MAX}. {\Dblue Specifically, in the real datasets Cora, Citeseer, Pubmed, and Coauthor-Physics, the standard deviation of \textsf{ALL} is 2.22, 2.95, 2.96, and 43.29 times higher than that of \textsf{Aug-MAX}, respectively.} This suggests that the imbalance in node usage frequency caused by using all cliques results in biased representation learning and unbalanced information propagation, leading to lower predictive performance compared to \textsf{Aug-MAX}. This serves as the primary motivation for introducing \textsf{Aug-MAX}, which offers a systematic approach to selecting cliques for higher-order representation learning.
\begin{figure*}[!t]
\centering
\includegraphics[width=\textwidth]{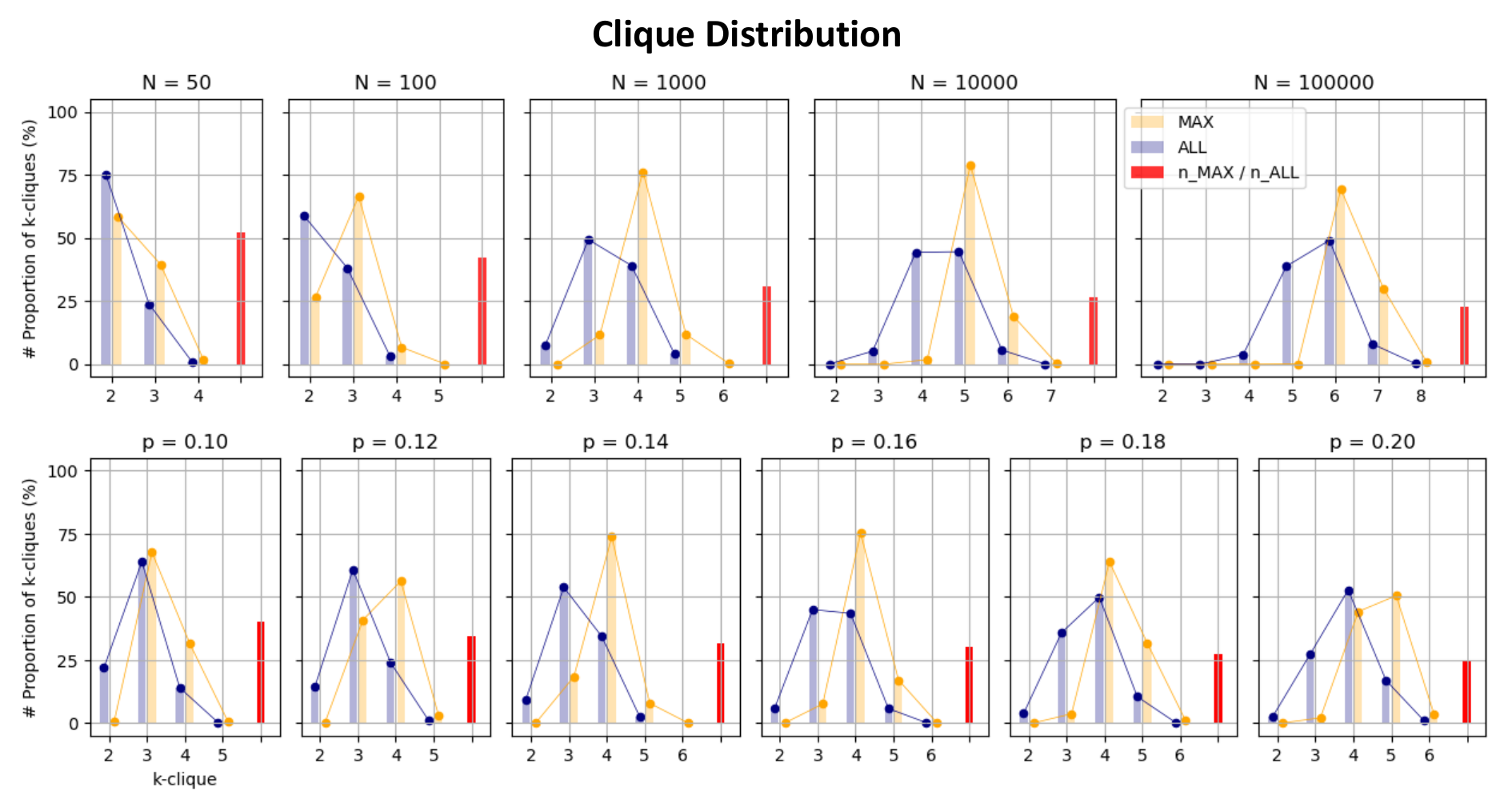}
\caption{
Clique size distributions in the PPM model with varying number of nodes and intra-class connection probability $p$. The upper panels illustrate the proportion of clique sizes for \textsf{ALL} (navy) and \textsf{MAX} (orange) under the setting $p=0.15$, $q=0.015$, and $[N,N,N]$, where the number of nodes is $N$ for each of the three labels. The lower panels show the clique size proportions for \textsf{ALL} and \textsf{MAX} as $p$ increases under the fixed setting $[1000,1000,1000]$, with $q$ set to $p/10$. Red bars indicate the ratio of the number of \textsf{MAX} cliques to the total number of \textsf{ALL} cliques. }
\label{figure:dist}
\end{figure*}
Third, although PPM does not perfectly replicate real datasets, it allows us to infer the types and numbers of cliques used for training in networks with a large number of nodes or edges. According to the expected clique count formulation of \textsf{ALL} and \textsf{MAX} in the PPM synthetic graphs derived in Appendix \ref{appendix}, as the number of nodes increases or the node connection probability grows, the proportion of higher-order cliques in the total set of cliques increases, while the ratio $\tfrac{|Q|}{|K|} = \tfrac{\sum_{i=2}^M |Q_k|}{\sum_{i=2}^M |K_k|}$ decreases (Figure \ref{figure:dist}). Notably, as the network size grows, in \textsf{MAX} and \textsf{Aug-MAX} strategies, only a small fraction of small-sized cliques participate in training.


{\Dblue Fourth, integrating \textsf{Aug-MAX} with 7 GNN models results in average accuracy gains of 2.14\%, 1.28\%, 0.56\%, and 0.25\% over the raw models for Cora, Citeseer, Pubmed, and Coauthor-Physics, respectively.} The objective function \eqref{objective3} is designed to encourage elements of higher-order cliques to share similar probability distributions, based on the assumption that densely interconnected nodes tend to have the same label. However, when the proportion of inter-cluster connections is high so that a large number of higher-order cliques consist of nodes from different labels, the proposed objective function struggles to achieve high performance. In the Cora, Citeseer, and Pubmed datasets, the proportion of inter-cluster cliques among 3-cliques and higher is 16.4\%, 18.8\%, and 22.6\%, respectively, with Citeseer and Pubmed exhibiting particularly high inter-cluster proportions. While the \textsf{Aug-MAX} strategy notably reduces the standard deviation in node participation frequency during training, a high inter-cluster ratio may pose a constraint on achieving additional performance gains.

Fifth, the \textsf{Aug-MAX} strategy incorporates some non-maximal cliques into \textsf{MAX} while significantly reducing the total number of cliques used for training compared to \textsf{ALL}. PPM simulations indicate that as the network size increases, the proportion of \textsf{MAX} to \textsf{ALL} further decreases. However, the computational complexity of \textsf{ALL} and \textsf{MAX} remains $O(\sum_{k=2}^M |K_k | k^l)$ and $O(\sum_{k=2}^M |Q_k | k^l)$, respectively, where $l$ is the number of labels. Despite the reduction in the total number of cliques, the $k^l$ term remains unchanged. Consequently, when the number of labels is large and the size of cliques grows, computational complexity remains a significant challenge. Additionally, since the polynomial coefficient $C_\tta$ in \eqref{objective}, \eqref{objective2} and \eqref{objective3} is computed differently for each $k$, it is difficult to perform computations on cliques of different sizes using a single method. This inherent complexity prevents a straightforward parallelization of the proposed algorithm. Addressing computational complexity and parallelization challenges is essential for enhancing the generalization and efficiency of the higher-order polyhedral-based objective function employed in this study, highlighting it as a key avenue for future research.

{\Dblue 
Sixth, while \textsf{Aug-MAX} effectively reduces redundancy compared to the \textsf{ALL} strategy, it does not fully resolve the imbalance in node participation. Certain nodes may still appear excessively across multiple maximal cliques due to overlap, leading to biased training. A promising future direction would be to integrate overlapping maximal cliques into unified structures, potentially through techniques such as link prediction, thereby reducing redundancy more fundamentally and improving computational efficiency. At the same time, we acknowledge that our framework relies on the enumeration of higher-order cliques, and for very large-scale graphs, this task is NP-hard and may become a computational bottleneck. This limitation is not unique to our method but is a common challenge faced by all clique-based, higher-order analysis methodologies.
}

\section{Conclusion}\label{conclusion}

Utilizing higher-order cliques to capture complex relational patterns between nodes or collective node patterns in network analysis is a natural approach. Previous studies \cite{koo2025node} have proposed a higher-order polyhedral-based objective function for the semi-supervised node classification task, where each node has a probability distribution, and training aims to make the distributions within each clique more similar across the network. However, training on all cliques increases computational complexity and raises the risk of overfitting. Additionally, nodes with high connectivity tend to appear in multiple cliques, leading to excessive participation in training and causing an over-representation issue.

To address these challenges, we propose the augmented maximal clique strategy (\textsf{Aug-MAX}). This strategy selectively incorporates non-maximal cliques into the maximal clique set, significantly reducing the total number of cliques used for training while also minimizing the standard deviation of node participation frequency compared to strategies that use all cliques. To validate the proposed strategy, we conduct experiments on balanced and imbalanced networks generated by PPM and further evaluate its effectiveness by integrating it with GNN-based models on real citation datasets.

Experimental results demonstrate that \textsf{Aug-MAX} significantly reduces the number of cliques used during training compared to approaches that utilize all cliques (\textsf{ALL}), indicating a substantially lower computational complexity for \textsf{Aug-MAX}. Additionally, by lowering the standard deviation of node participation frequency compared to \textsf{ALL} and \textsf{MAX}, it mitigates biased representation learning and improves classification accuracy.

Despite the efficiency and accuracy of the proposed strategy, the current algorithm faces limitations in parallelization due to the distinct training mechanisms required for cliques of varying sizes. Additionally, in networks with a high proportion of structurally diverse nodes, particularly those involving numerous higher-order cliques with mixed labels, computational complexity can become a concern. Overcoming these challenges remains an important direction for future research.

\appendix

\section{Derivation of the expected number of maximal cliques}\label{appendix}

We derive explicit formulas for the expected number of all cliques and maximal cliques in the PPM model, considering arbitrary node counts across labels and arbitrary intra- and inter connection probabilities, denoted by $p$ and $q$, respectively. Let the label index be $I=\{1,2,...,l\}$, and the number of nodes with label $i$ be $N_i$ for each $i \in I$, so that $N = \sum_{i=1}^l N_i$. Each $k \in \N$ will represent the size of a clique. Let $\pi(k)$ denote the set of partitions of $k$. We list the partition of $k$ in increasing order. For example, $\pi(4) = \{(1,1,1,1),(1,1,2),(1,3),(2,2),(4)\}$, thus $|\pi(4)| = 5$. Each $\eta \in \pi(k)$ represents the label composition in a $k$-clique. For example, $\eta = (1,1,2)$ tells us that the 4-clique under consideration has 3 distinct labels assigned to its 4 nodes, where two nodes have the same label. Let $|\eta|$ denote the number of entries in $\eta$ (i.e., the number of distinct labels in a clique), such that $\eta= (\eta_1,...,\eta_{|\eta|})$. Let $\eta_+ = \sum_{i=1}^{|\eta|} \eta_i$, and define $\eta ! = \eta_1 ! \eta_2 ! \dots \eta_{|\eta|} !$. For example, if $\eta=(2,2,3)$, then $|\eta| = 3$, $\eta_+=7$ and $\eta!=2!2!3!=24$. For $L,M \in \N$, let $S(L,M)$ be the set containing all possible permutations of $M$ distinct numbers selected from $\{1,2,...,L\}$. For example, $S(4,2) = \{(1,2),(1,3),(1,4),(2,1),(2,3),(2,4),(3,1),(3,2),(3,4),\\ (4,1),(4,2),(4,3)\}$. Let $S(L,M)= \emptyset$ if $L < M$.

\begin{proposition}\label{cliqueformula}
For each $k \ge 2$, let $\E[K_k]$ and $\E[Q_k]$ represent the expected number of the $k$-cliques, and of the maximal $k$-cliques in the PPM, respectively. Then it holds:
\begin{align}
&\E[K_k]= \sum_{\eta \in \pi(k)} P(\eta) \cdot \frac{1}{\eta !}\sum_{(j_1,...,j_{|\eta|}) = \xi \in S(l, |\eta|)} A(\eta, \xi), \label{A1} \\
&\E[Q_k] \nn \\
&=  \sum_{\eta \in \pi(k)} \frac{P(\eta)}{\eta !} \sum_{(j_1,...,j_{|\eta|}) = \xi \in S(l, |\eta|)} A(\eta, \xi) B_1(\eta, \xi) B_2(\eta, \xi),\label{A2}
\end{align}
where (assuming ${n \choose m} =0$ if $n < m$)
\begin{align*}
P(\eta) &= p^{\sum_{m=1}^{|\eta|} {\eta_m \choose 2}} q^{{\eta_+ \choose 2} - \sum_{m=1}^{|\eta|} {\eta_m \choose 2}}, \\
A(\eta, \xi) &= \prod_{m=1}^{|\eta|} {N_{j_m} \choose \eta_m},  \\
B_{1}(\eta, \xi) &= \prod_{m=1}^{|\eta |} (1 - p^{\eta_m} q^{\eta_+ - \eta_m})^{N_{j_m} - \eta_m},\\
B_{2}(\eta, \xi) &= \prod_{i \in I \setminus \{j_1,...,j_{|\xi|}\} } (1 - q^{\eta_+})^{N_i}.
\end{align*}
Hence, $\E[K] = \sum_{k=2}^N \E[K_k]$ and $\E[Q]= \sum_{k=2}^N  \E[Q_k]$ represent the expected number of all cliques, and of the maximal cliques, respectively, in the PPM.
\end{proposition}

\begin{proof}
It suffices to interpret the terms \(P,A,B_1\), and \(B_2\). Given \( \eta \in \pi(k) \) (thus \( k = \eta_+ \)), \( P(\eta) \) represents the probability of the existence of a \( k \)-clique in the graph whose node label counting is given by \( \eta \). This is because the total number of edges in the \( k \)-clique connecting nodes with the same label is \( \sum_{m=1}^{|\eta|} {\eta_m \choose 2} \), and hence, the number of edges connecting nodes with different labels is \( {k \choose 2} - \sum_{m=1}^{|\eta|} {\eta_m \choose 2} \). Meanwhile, $ \displaystyle \frac{1}{\eta !}\sum_{(j_1,...,j_{|\eta|}) = \xi \in S(l, |\eta|)} A(\eta, \xi)$ represents the total number of the set of \( k \) nodes whose label counting is given by \( \eta \). This yields \eqref{A1}. To derive \eqref{A2}, let us first consider the simpler case where the probability of any two nodes being connected is $p$. The expected number of $k$-cliques is then  ${N \choose k} p^{{k \choose 2}}$. Now observe that a $k$-clique is a maximal clique if and only if there exists no outside node that connects to all of the $k$ nodes in the clique. This observation implies that the expected number of maximal $k$-clique in this simple case is ${N \choose k} p^{{k \choose 2}} (1-p^k)^{N-k}$. By generalizing the term \( (1-p^k)^{N-k} \) for two parameters \( (p,q) \), we can derive \( B_1, B_2 \). Given a clique \( \sigma \) with \( \eta \) its node label counts, we divide the label index \( I \) into two parts: labels used to label the nodes in \( \sigma \) (denoted as \( I_1 \)) and labels not used (denoted as \( I_2 \)). Then \( B_j(\eta, \xi) \), \( j=1,2 \), represents the probability that any node outside \( \sigma \) labeled in \( I_j \) does not connect to all of the nodes in \( \sigma \). This completes the proof.
\end{proof}

\section*{Acknowledgments}
Eunho Koo acknowledges the support of the National Reserach Foundation of Korea (NRF) grant funded by the Korea government (Grant No. RS-2025-02363044).

\bibliographystyle{IEEEtran}
\bibliography{MaximalClique}


%


\end{document}